\documentclass[aps,prl,twocolumn,showpacs]{revtex4}
\usepackage{amsfonts}
\usepackage{mathrsfs}
\begin{document}
\title{Spin and orbital angular momentum in gauge theories:\\
Nucleon spin structure and multipole radiation revisited}
\author{Xiang-Song Chen,$^{1,2,}$\footnote{Electronic address:
cxs@scu.edu.cn}
Xiao-Fu L\"{u},$^1$ Wei-Min Sun,$^2$ Fan Wang,$^2$ and T.
Goldman$^{3,}$\footnote{Electronic address:
t.goldman@post.harvard.edu}} \affiliation{$^1$Department of Physics,
Sichuan University, Chengdu 610064, China\\
$^2$Department of Physics, Nanjing University, CPNPC, Nanjing
210093,
China\\
$^3$Theoretical Division, Los Alamos National Laboratory, Los
Alamos, NM 87545, USA}
\date{Phys. Rev. Lett. {\bf 100}, 232012, 12 June 2008}

\begin{abstract}
We address and solve the long-standing gauge-invariance problem of
the nucleon spin structure. Explicitly gauge-invariant spin and
orbital angular momentum operators of quarks and gluons are
obtained. This was previously thought to be an impossible task, and
opens a more promising avenue towards the understanding of the
nucleon spin. Our research also justifies the traditional use of the
canonical, gauge-dependent angular momentum operators of photons and
electrons in the multipole-radiation analysis and labeling of atomic
states; and sheds much light on the related energy-momentum problem
in gauge theories, especially in connection with the nucleon
momentum. 
\end{abstract}
\maketitle

\textit{The dilemma in separating the nucleon spin} --- As a
composite particle, the nucleon naturally obtains its spin from the
spin and orbital motion of its constituents: quarks and gluons. From
a theoretical point of view, the first task in studying the nucleon
spin structure is to find out the appropriate operators for the spin
and orbital angular momentum of the quark and gluon fields. Given
these operators, one can then study their matrix elements in a
polarized nucleon state, and investigate how these matrix elements
can be related to experimental measurements. Disturbingly and
surprisingly, after 20 years of extensive discussions of the nucleon
spin structure~\cite{Bass07,Herm07,Stra07,Bass05,EMC}, this first
task has never been done, and has even largely eluded the attention
of the community.

At first thought, it seems an elementary exercise to derive the
quark and gluon angular momentum operators. From the Lagrangian $
\mathscr{L}=-\frac 14F^a_{\mu\nu}F^{a\mu\nu}+ \bar\psi( i\gamma^\mu
\mathbb{D}_\mu -m )\psi $, where $\mathbb{D}_\mu =\partial_\mu+ig
\mathbb{A}_\mu$ and $\mathbb{A}_\mu\equiv A^a_\mu T^a$ (with $T^a$
the generators of the color SU(3) group), one can promptly follow
N\"{o}ther's theorem to write down the canonical expression of the
conserved QCD angular momentum:
\begin{eqnarray}
\vec J_{QCD} &=& \int d^3 x \psi ^\dagger \frac 12 \vec \Sigma \psi
+ \int d^3x \psi ^\dagger \vec x \times\frac 1i \vec \nabla \psi
\nonumber \\
 &+&\int d^3x \vec E^a\times \vec A^a+
\int d^3x E^{ai}\vec x\times \vec \nabla A^{ai} \nonumber \\
&\equiv& \vec S_q +\vec L_q +\vec S_g +\vec L_g, \label{QCD1}
\end{eqnarray}
and readily identify the four terms here as the quark spin ($\vec
\Sigma={\rm diag.} \left(\vec \sigma, \vec \sigma \right)$ and $\vec
\Sigma \times \vec \Sigma =i\vec \Sigma$), quark orbital angular
momentum, gluon spin, and gluon orbital angular momentum,
respectively. However, except for the quark spin, all of the other
three terms are gauge dependent, thus have obscure physical
meanings. In this regard, it should be noted that the total angular
momentum is nonetheless gauge invariant (as it must be). This can be
seen from an alternative, explicitly gauge invariant expression
\cite{Ji97,Chen97}:
\begin{eqnarray}
\vec J_{QCD} &=& \int d^3 x \psi ^\dagger \frac 12 \vec \Sigma \psi
+ \int d^3x \psi ^\dagger \vec x \times\frac 1i \vec \mathbb{D}
\psi\nonumber\\
&+&\int d^3x \vec x\times (\vec E^a\times\vec B^a) \nonumber \\
&\equiv& \vec S_q +\vec L'_q +\vec J'_g.  \label{QCD2}
\end{eqnarray}
This is obtained from Eq.(\ref{QCD1}) by adding a surface term,
\begin{equation}
 \int d^3 x\vec\nabla \cdot [\vec E^a(\vec A^a\times \vec x)],
 \label{sur}
\end{equation}
which vanishes after integration. Since all of the terms in
Eq.(\ref{QCD2}) are separately gauge invariant, it may seem
appropriate to identify $\vec L'_q$ as the quark orbital angular
momentum, and $\vec J'_g$ as the total gluon angular momentum.
However, a further decomposition of $\vec J'_g$ into gauge invariant
gluon spin and orbital parts is lacking. Moreover, neither $\vec
L'_q$ nor $\vec J'_g$ obeys the fundamental angular momentum
algebra, $\vec J \times \vec J =i \vec J$ (although $\vec J'_g$ does
when the quark field is absent); hence they cannot be the relevant
rotation generators \cite{Chen97}. It has long been assumed by the
community that the reconciliation of gauge invariance and the
angular momentum algebra is not possible, and that gauge invariant,
local gluon spin and orbital angular momentum operators do not exist
\cite{Bass05}.

\textit{The QED problem revisited} --- Since QED is also a gauge
theory, the problems above first emerged there. In fact, by simply
dropping the color indices, Eqs.~(\ref{QCD1}) and (\ref{QCD2})
become exactly the expressions for the electron and photon angular
momenta, which we denote as:
\begin{eqnarray}
\vec J_{QED} &=& \vec S_e +\vec L_e +\vec S_\gamma +\vec L_\gamma
\label{QED1}\\
 &=&\vec S_e +\vec L'_e +\vec J'_\gamma.  \label{QED2}
\end{eqnarray}
Eq.(\ref{QED2}) is obtained from Eq.(\ref{QED1}) by adding the same
surface term as in (\ref{sur}), but without the color indices.

Similarly to the situation in QCD, neither Eq.(\ref{QED1}) nor
Eq.(\ref{QED2}) is fully satisfactory: On the one hand, the
canonical angular momentum operators in Eq.(\ref{QED1}) are what
people use familiarly in discussing polarized atomic states and
radiation, but the gauge dependence of these operators leads to an
uneasy concern about many calculations. As one example, the labeling
of atomic states, which uses eigenvalues of the electron orbital
angular momentum operator $\vec L_e=\int d^3x \psi ^\dagger \vec x
\times\frac 1i \vec \nabla \psi$, seems gauge dependent! For another
example, the multipole-radiation analysis, which employs the
multipole-field wavefunction constructed with photon spin and
orbital angular momentum operators, seems again gauge dependent! On
the other hand, the gauge-invariant operators in Eq.(\ref{QED2}),
$\vec L'_e$, and $\vec J'_\gamma$, are not appropriate for
constructing angular momentum eigenstates (because, as we remarked
above, they are not angular momentum operators at all), and do not
separate photon spin from photon orbital angular momentum. It is
stated in common textbooks that gauge invariance prohibits the
separation of photon angular momentum into spin and orbital
contributions~\cite{Jauc55,Bere82}, yet both photon spin and orbital
angular momentum have been measured separately by experiments
\cite{Beth36,Enk07,Ande06,Alex06,Marr06,Leac02,Alle92}.

Despite the gauge dependence of Eq.(\ref{QED1}), all QED angular
momentum calculations based on it seem to agree well with
experiments. It is therefore hard to believe that all those
discussions, including the whole multipole-radiation analysis and
labeling of atomic, nuclear, and hadronic states, are meaningless.
Enlightened by earlier clarifications \cite{Enk94,Calv06}, we find
that there exists indeed a satisfactory and decisive answer for the
question of spin and orbital angular momentum in QED:
\begin{eqnarray}
\vec J_{QED} &=& \int d^3 x \psi ^\dagger \frac 12 \vec \Sigma \psi
+ \int d^3x \psi ^\dagger \vec x \times\frac 1i \vec D_{pure} \psi
\nonumber \\
 &+&\int d^3x \vec E \times \vec A_{phys}+ \int d^3x
E^i\vec
x\times \vec \nabla A_{phys}^i \nonumber \\
&\equiv& \vec S_e +\vec L''_e +\vec S''_\gamma +\vec L''_\gamma
\label{QED3}.
\end{eqnarray}
Here, $\vec D_{pure} \equiv\vec \nabla -ie\vec A_{pure}$, $\vec
A_{pure}+\vec A_{phys}\equiv \vec A$ and the two parts are defined
via:
\begin{eqnarray}
\vec \nabla\cdot \vec A_{phys} =0,\label{AphysE}\\
\vec \nabla \times \vec A_{pure}=0. \label{ApureE}
\end{eqnarray}
These are nothing but the transverse and longitudinal components of
the vector potential $\vec A$. The subscripts used here are intended
to make the physical ({\it vs.} pure-gauge) content clear, and to
prepare for the generalization to QCD. With the boundary condition
that $\vec A$, $\vec A_{pure}$, and $\vec A_{phys}$ all vanish at
spatial infinity, Eqs. (\ref{AphysE}) and (\ref{ApureE}) prescribe a
unique decomposition of $\vec A$ into $\vec A_{pure}$ and $\vec
A_{phys}$, and dictate their distinct gauge transformation
properties:
\begin{eqnarray}
\vec A_{pure}&\rightarrow & \vec A'_{pure}= \vec A_{pure} +\vec
\nabla \Lambda, \label{ApureE'}\\
\vec A_{phys}&\rightarrow &\vec A'_{phys}= \vec A_{phys},
\label{AphysE'}
\end{eqnarray}
under a gauge transformation $\Lambda$. Eqs. (\ref{ApureE}) and
(\ref{ApureE'}) tell us that, in QED, $\vec A_{pure}$ is a
pure-gauge field in all gauges, and that it transforms in the same
manner as does the full vector field: $\vec A\rightarrow \vec A'=
\vec A +\vec \nabla \Lambda $. On the other hand, the transverse
field $\vec A_{phys}$ is unaffected by gauge transformations, and so
can be regarded as the ``physical'' part of $\vec A$.

Eq.(\ref{QED3}) is obtained from Eq.(\ref{QED1}) by adding another
surface term,
\begin{equation}
 \int d^3 x \vec\nabla \cdot [\vec E(\vec A_{pure}\times \vec
 x)]. \label{surpure}
\end{equation}

Now we have all of the elements needed to explain how
Eq.(\ref{QED3}) gives the correct expressions for the spin and
orbital angular momenta of electrons and photons, including their
densities. First of all, the total $\vec J_{QED}$ given by
Eq.(\ref{QED3}) equals that in Eqs.(\ref{QED1}) and (\ref{QED2}),
for they merely differ by surface terms. Second, the gauge
transformation properties of $\vec A_{pure}$ and $\vec A_{phys}$
show that each {\it density} term in Eq.(\ref{QED3}) is separately
gauge invariant (and hence, so is the integrated operator). Third,
like the canonical $\vec L_e$, the gauge invariant $\vec L''_e$
satisfies the angular momentum algebra $\vec J\times\vec J=i\vec J$.
This is due to the property of $\vec A_{pure}$ in Eq.(\ref{ApureE}).
Finally, we note that, in Coulomb gauge, $\vec\nabla \cdot \vec
A=0$, so the longitudinal (pure-gauge) field $\vec A_{pure}$
vanishes; thus all quantities in Eq.(\ref{QED3}) coincide with their
canonical counterparts in Eq.(\ref{QED1}). This observation is of
vital importance. It reveals that the gauge invariant quantities in
Eq.(\ref{QED3}) can all be conveniently computed via the canonical
operators in Coulomb gauge. This is actually what people
(implicitly) do in studying atomic and electromagnetic angular
momenta (such as in multipole radiation), including the recent
measurements of the photon orbital angular
momentum~\cite{Enk07,Ande06,Alex06,Marr06,Leac02,Alle92}. It is thus
natural that these studies always obtain reasonable results.

\textit{Hindsight for QED and solution for QCD} --- After confirming
that Eq.(\ref{QED3}) is indeed the correct and satisfactory answer
for angular momenta in QED, we can observe something about it in
hindsight: The form of Eq.(\ref{QED3}) could have been guessed by
reasonable physical considerations: The photon angular momentum
should contain only the ``physical'' part of the gauge field, which
should, nevertheless, not appear in the expression for the electron
orbital angular momentum $\vec L''_e$. The latter should thus only
include the non-physical $\vec A_{pure}$, so as to cancel the
equally non-physical phase dependence of the electron field, and
keep the whole $\vec L''_e$ gauge invariant. From this hindsight for
QED, it is natural to expect that the correct, gauge invariant
expressions of QCD angular momenta should be:
\begin{eqnarray}
\vec J_{QCD} &=& \int d^3 x \psi ^\dagger \frac 12 \vec \Sigma \psi
+ \int d^3x \psi ^\dagger \vec x \times\frac 1i \vec
\mathbb{D}_{pure} \psi
\nonumber \\
 &+& \int d^3x \vec E^a \times \vec A^a_{phys}+ \int
d^3x
E^{ai}\vec x\times \vec \nabla A^{ai}_{phys} \nonumber \\
&\equiv& \vec S_q +\vec L''_q +\vec S''_g +\vec L''_g \label{QCD3},
\end{eqnarray}
where $\vec \mathbb{D}_{pure} \equiv\vec \nabla -ig\vec
\mathbb{A}_{pure}$ and $\vec \mathbb{A}_{pure}\equiv \vec
A^a_{pure}T^a$. The essential task remaining now is to properly
define the pure-gauge field $\vec \mathbb{A}_{pure}$ and the
``physical'' field $\vec \mathbb{A}_{phys}\equiv\vec A^a_{phys}T^a$
so that they have the desired gauge transformation properties, and
to prove that the sum of the four terms in Eq.(\ref{QCD3}) equals
that in Eqs.(\ref{QCD1}) and (\ref{QCD2}). This, however, turns out
to be non-trivial.

The parallel construction of Eqs.(\ref{AphysE}) and (\ref{ApureE})
obviously does not work in QCD: For one thing, $\vec
\mathbb{A}_{pure}$ defined by $\vec \nabla \times \vec
\mathbb{A}_{pure}=0$ is not a pure-gauge term in QCD; for another,
$\vec \nabla \cdot \vec \mathbb{A}_{phys}=0$ and $\vec \nabla \times
\vec \mathbb{A}_{pure}=0$ are not invariant under the SU(3) gauge
transformation:
\begin{equation}
\mathbb{A}_\mu\rightarrow \mathbb{A}'_\mu =U\mathbb{A}_\mu U^\dagger
-\frac ig U\partial_\mu U^\dagger.
\end{equation}

To make $\vec \mathbb{A}_{pure}$ a pure-gauge term in QCD, we
require, instead of Eq. (\ref{ApureE}), that
\begin{equation}
\vec \mathbb{D}_{pure}\times \vec \mathbb{A}_{pure} =\vec \nabla
\times \vec \mathbb{A}_{pure} -ig \vec \mathbb{A}_{pure}\times \vec
\mathbb{A}_{pure}=0. \label{ApureC}
\end{equation}
This provides two independent equations for $\vec
\mathbb{A}_{pure}$. We still need a third equation that plays the
same role as Eq.(\ref{AphysE}) does in QED, so that $\vec
\mathbb{A}_{phys}$ and $\vec \mathbb{A}_{pure}$ have the required
transformation properties:
\begin{eqnarray}
\vec \mathbb{A}_{pure}&\rightarrow & \vec \mathbb{A}'_{pure}= U\vec
\mathbb{A}_{pure} U^\dagger +\frac ig U\vec
\nabla U^\dagger, \label{ApureC'}\\
\vec \mathbb{A}_{phys}&\rightarrow &\vec \mathbb{A}'_{phys}= U\vec
\mathbb{A}_{phys} U^\dagger. \label{AphysC'}
\end{eqnarray}

To seek this third equation, we proceed inversely by applying these
transformations to examine the gauge invariance of each operator in
Eq.(\ref{QCD3}). The reason why this is possible will be clear
shortly below.

The quark orbital angular momentum $\vec L''_q$ provides no further
constraints. Eqs.(\ref{ApureC}) and (\ref{ApureC'}) guarantee its
gauge invariance, as well as the correct angular momentum algebra
$\vec L''_q\times \vec L''_q=i\vec L''_q$. The gluon spin $\vec
S''_g$ provides no further constraints either. Eq.(\ref{AphysC'})
tells us that it is gauge invariant. However, the situation for the
gluon orbital angular momentum $\vec L''_g$ is different: Unlike in
QED, $\vec \mathbb{A}_{phys}$ here is gauge covariant instead of
invariant, which leads to the gauge transformation of $\vec L''_g$:
\begin{eqnarray}
&&E^{ai}\vec x\times \vec \nabla
A^{ai}_{phys}=2\textrm{Tr}\{\mathbb{E}^i\vec x\times \vec \nabla
\mathbb{A}_{phys}^i \}\nonumber \\ &\rightarrow&
2\textrm{Tr}\{U\mathbb{E}^iU^\dagger\vec x\times \vec \nabla (U
\mathbb{A}_{phys}^i U^\dagger
)\}\nonumber \\
&=& 2\textrm{Tr}\{\mathbb{E}^i\vec x\times \vec \nabla
\mathbb{A}_{phys}^i
\} \nonumber \\
& +& 2 \textrm{Tr}\{\vec x\times U^\dagger (\vec\nabla U)(\vec
\mathbb{A}_{phys}\cdot \vec \mathbb{E}-\vec \mathbb{E}\cdot\vec
\mathbb{A}_{phys}) \},
\end{eqnarray}
where $\vec \mathbb{E}\equiv \vec E^aT^a$. Hence, to make $\vec
L''_g$ invariant under arbitrary gauge transformations, we have to
set
\begin{equation}
[\vec \mathbb{A}_{phys}, \vec \mathbb{E} ]\equiv\vec
\mathbb{A}_{phys}\cdot \vec \mathbb{E}-\vec \mathbb{E}\cdot\vec
\mathbb{A}_{phys}=0. \label{AphysC}
\end{equation}
This is the third equation that we seek. The remaining task is to
cross-check the consistency of whether or not Eqs.(\ref{ApureC}) and
(\ref{AphysC}) dictate the transformation properties in
Eqs.(\ref{ApureC'}) and (\ref{AphysC'}).

Before making this cross-check, we first make another vital check,
namely, whether the definitions of $\vec \mathbb{A}_{pure}$ and
$\vec \mathbb{A}_{phys}$ by Eqs.(\ref{ApureC}) and (\ref{AphysC})
ensure that the total angular momentum in Eq.(\ref{QCD3}) is to
equal that in Eqs.(\ref{QCD1}) and (\ref{QCD2}). Since no more
tricks are available, the answer must be positive or our entire
approach will founder. A slightly lengthy but straightforward
calculation shows that the answer is indeed positive: With the
definitions in Eqs.(\ref{ApureC}) and (\ref{AphysC}),
Eq.(\ref{QCD3}) can be obtained from Eq.(\ref{QCD1}) by adding a
surface term similar to (\ref{surpure}) for QED:
\begin{equation}
 \int d^3 x \vec\nabla \cdot [\vec E^a(\vec A_{pure}^a\times \vec
 x)].
\end{equation}

As to the cross-check, we note that $\vec \mathbb{A}'_{pure}$ and
$\vec \mathbb{A}'_{phys}$ given by Eqs.(\ref{ApureC'}) and
(\ref{AphysC'}) are solutions of
\begin{eqnarray}
\vec \mathbb{D}'_{pure}\times \vec \mathbb{A}'_{pure}&=&0,
\label{ApureC"}\\
\left[\vec \mathbb{A}'_{phys}, \vec \mathbb{E}'\right] &=&0 ,
\label{AphysC"}
\end{eqnarray}
where $\vec \mathbb{D}'_{pure}\equiv \vec \nabla-ig
\mathbb{A}'_{pure}$ and $\vec \mathbb{E}'=U\vec \mathbb{E}
U^\dagger$. The remaining question is whether Eqs.(\ref{ApureC"})
and (\ref{AphysC"}) have any other solution than that given by
Eqs.(\ref{ApureC'}) and (\ref{AphysC'}). This is equivalent to
asking whether Eqs.(\ref{ApureC}) and (\ref{AphysC}) uniquely
determine the decomposition of $\vec \mathbb{A}$ into $\vec
\mathbb{A}_{pure}$ and $\vec \mathbb{A}_{phys}$, or, essentially,
whether the constraint, $ [\vec \mathbb{A}, \vec \mathbb{E} ]=0$,
fixes the gauge completely. This is a tricky question, for, unlike
in QED, many gauges in QCD suffer from topological complexity such
as Gribov copies~\cite{Grib78}. Fortunately, such complexity does
not bother us here: If supplementary conditions are needed to
restrict the solutions of Eqs.(\ref{ApureC"}) and (\ref{AphysC"}) to
that given by Eqs.(\ref{ApureC'}) and (\ref{AphysC'}), they can
simply be added, without affecting the equivalence of
Eq.(\ref{QCD3}) with Eqs.(\ref{QCD1}) and (\ref{QCD2}), and without
affecting the gauge invariance of the angular momentum operators we
constructed; because these properties rely only on
Eqs.(\ref{ApureC}), (\ref{ApureC'}), (\ref{AphysC'}), and
(\ref{AphysC}).

\textit{Remarks and discussion} --- 1) We have noted that for QED in
Coulomb gauge, $\vec \nabla \cdot \vec A=0$, Eq.(\ref{QED3})
coincides with Eq.(\ref{QED1}). Similarly, for QCD, in the gauge
$[\vec \mathbb{A}, \vec \mathbb{E} ]=0$ (together with possible
supplementary conditions to completely fix the gauge),
Eq.(\ref{QCD3}) coincides with Eq.(\ref{QCD1}). Namely, in actual
calculations, QCD shares the same nice feature as in QED that the
gauge-invariant, physically meaningful angular momenta can be
conveniently computed via their canonical, gauge-dependent
counterparts in a ``physical'' gauge in which the pure-gauge
component vanishes. From the QCD equation of motion, $\vec \nabla
\cdot \vec \mathbb{E}=ig [\vec \mathbb{A}, \vec \mathbb{E}
]+g\psi^\dagger T^a\psi T^a$, we see that the gauge $[\vec
\mathbb{A}, \vec \mathbb{E} ]=0$ says essentially that the
(gauge-dependent) color charge carried by gluons vanishes. So $[\vec
\mathbb{A}, \vec \mathbb{E} ]=0$ has the sense of a ``generalized''
Coulomb gauge, for it leads to an equation of motion $\vec \nabla
\cdot \vec E^a=g\psi^\dagger T^a\psi$, similar to Gauss' law in QED.

2) Our construction guarantees that all angular momentum operators
transform properly under spatial translation and rotation. To figure
out how they transform under boost, we need to carry out the
canonical quantization procedure (preferably in the ``physical''
gauge in which the pure-gauge terms vanish), and compute the
commutators of the angular momentum operators with the
interaction-involving boost generators. This non-trivial task will
be our next work.

3) In the literature, there have been various discussions about
decomposing the Yang-Mills field into several components
representing different degrees of freedom, based mainly on
group-theoretical considerations \cite{Fadd07,Marn82,Duan79}. It
would be very interesting to investigate how these decompositions
are related to ours, which is dictated by the requirement of a
physically meaningful angular momentum expression.

4) The so-called gluon polarization $\Delta G$ being measured at
several facilities~\cite{Stra07} is related to $\vec S_g$ in
Eq.(\ref{QCD1}) in the temporal gauge in the infinite-momentum frame
of the nucleon~\cite{Jaff96}. From our discussion, $\Delta G$ is not
the gauge invariant gluon spin $S''_g$ that we construct here.

5) Beth made a direct measurement of the photon spin over 70 years
ago~\cite{Beth36}. Detection and manipulation of the photon orbital
angular momentum have also been carried out recently, and have
become a hot topic due to their potential application in quantum
information processing
\cite{Enk07,Ande06,Alex06,Marr06,Leac02,Alle92}. These measurements
can be straightforwardly interpreted with the operators in
Eq.(\ref{QED3}), via its equivalence to Eq.(\ref{QED1}) in the
Coulomb gauge. This should encourage the investigation of the
picture of the nucleon spin in terms of the gauge-invariant,
physically meaningful decomposition in Eq.(\ref{QCD3}), which is
completely analogous to Eq.(\ref{QED3}) for QED. Experimentally, the
free-beam-based photon measurements can certainly not be extended to
gluons directly, and appropriate (even ingenious!) methods for
measuring $L''_q$, $S''_g$, and $L''_g$ will have to be invented;
but the clear physical meanings and explicit gauge invariance of
these quantities guarantee at least that there can be pertinent
theoretical calculations of them, especially in lattice QCD.

6) From the correct, gauge-invariant angular momentum expression in
Eq.(\ref{QED3}), we can read out the correct electromagnetic
momentum {\em density} to be $E^i\vec \nabla A^i_{phys}$, instead of
the renowned Poynting vector, $\vec E \times \vec B$. The latter
actually includes a spin current, and can be unambiguously
distinguished from the purely mechanical momentum $E^i\vec \nabla
A^i_{phys}$ by delicate measurement \cite{Chen07}. In QCD, $\vec E^a
\times \vec B^a$ leads to a picture that gluons carry half of the
nucleon momentum on the light-cone \cite{Sloa88}. This picture may
therefore need to be revised. Similarly to the situation for the
angular momentum, the momentum operators we propose transform
properly under spatial translation and rotation, and next, we will
study how they transform under boost by computing their commutators
with the boost generators via canonical quantization.

We thank C.D. Roberts, T. Lu, and S.J. Wang for helpful discussions.
This work is supported in part by the China NSF under grants
10475057 and 90503011, and in part by the U.S. DOE under contract
W-7405-ENG-36.

\end{document}